\documentclass{ifacconf}

\usepackage{graphicx}      
\usepackage{natbib}        

\usepackage{siunitx} 
\usepackage{xcolor} 
\usepackage{amsmath,amssymb,amsfonts,mathtools,tensor} 
\usepackage{blindtext} 

\usepackage{booktabs}
\usepackage{tabu}
\usepackage{multirow}

\usepackage{pgfplots}
\usepackage{varwidth}
\usetikzlibrary{arrows}

\usepackage{caption}
\usepackage{subcaption}

\newcommand{\m}[1]{\boldsymbol{#1}} 
\DeclareMathOperator*{\argmax}{arg\,max} 
\newcommand{\mat}[1]{\begin{bmatrix*}[r]#1 
\end{bmatrix*}}

\newcommand{\sgn}{\mathop{\mathrm{sgn}}} 

\newcommand\td[1]{} 
\newcommand\tdd[1]{} 
\newcommand\comm[1]{} 
\newcommand\commd[1]{} 
\newcommand\fsam[1]{} 
\newcommand\fsamd[1]{} 
\newcommand\fmich[1]{} 
\newcommand\fmichd[1]{} 

\allowdisplaybreaks

\usepackage{algorithm}
\usepackage{algpseudocode} 
\algnewcommand{\Initialize}[1]{%
	\State \textbf{initialize} {\raggedright #1}
}
\makeatletter
\algnewcommand{\Statey}[1]{\Statex \hskip\ALG@thistlm #1}
\makeatother
\algdef{SE}[DOWHILE]{Do}{doWhile}{\algorithmicdo}[1]{\algorithmicwhile\ #1}%

\begin{document}
	\begin{frontmatter}
		
		\title{Deep Decentralized Reinforcement Learning for Cooperative Control\thanksref{footnoteinfo2}}
		
		\thanks[footnoteinfo]{These authors contributed equally to this work.}
		\thanks[footnoteinfo2]{This work has been submitted to IFAC for possible publication.}
		
		\author{Florian K\"opf,\thanksref{footnoteinfo}} 
		\author{ Samuel Tesfazgi\thanksref{footnoteinfo},} 
		\author{Michael Flad and S\"oren Hohmann}
		
		\address{Institute of Control Systems, Karlsruhe Institute of Technology (KIT), 76131~Karlsruhe, Germany (e-mail: florian.koepf@kit.edu)}
		
		\begin{abstract}                
			In order to collaborate efficiently with unknown partners in cooperative control settings, adaptation of the partners based on online experience is required. The rather general and widely applicable control setting, where each cooperation partner might strive for individual goals while the control laws and objectives of the partners are unknown, entails various challenges such as the non-stationarity of the environment, the multi-agent credit assignment problem, the alter-exploration problem and the coordination problem. We propose new, modular deep decentralized Multi-Agent Reinforcement Learning mechanisms to account for these challenges. Therefore, our method uses a time-dependent prioritization of samples, incorporates a model of the system dynamics and utilizes variable, accountability-driven learning rates and simulated, artificial experiences in order to guide the learning process. The effectiveness of our method is demonstrated by means of a simulated, nonlinear cooperative control task.	
		\end{abstract}
		
		
	\end{frontmatter}
	
	\section{Introduction}\label{sec:introduction}
	In numerous control problems including highly-automated driving, robotics and manufacturing plants, several entities (e.g. machines and/or humans) are required to collaborate in order to achieve complex control objectives. Although the cooperating partners' goals usually do not completely contradict each other, the partners might have individual preferences.
	Suitable partners need to be flexible enough to account for the preferences of each other while representing their interests.
	We refer to this kind of setting as \textit{Cooperative Control} \citep{Kopf.2018}\footnote{Alternatively termed \textit{Mixed Cooperative-Competitive Control} \citep{Lowe.2017}.} in order to emphasize that partners need to cooperate with each other and make compromises when conflicts occur. However, this does not necessarily imply that they are facing a so-called \textit{fully cooperative} setting with a single global goal. Instead individual goals for the agents are allowed.
	Bearing the vision of future human-machine collaboration and plug-and-play machine-machine cooperation in mind, 
	we focus on the case where the partners do not know the others' control laws or objective functions and no explicit communication is used. This \textit{decentralized} setting requires the partners to constantly adapt to each other based on online experience. 
	
	Due to its generalization capabilities and major successes in the single-agent Reinforcement Learning (RL) setting, Multi-Agent Reinforcement Learning (MARL) has recently become the focus of increasing attention in order to solve Cooperative Control problems. Compared to single-agent RL, the multi-agent case is inherently more complex as agents directly or indirectly interact with each other and their common environment.	
	
	A major challenge occurring here is the \textit{non-stationarity} of the dynamics from the local perspective of each agent 
	which violates the Markov property that is commonly assumed in RL. 
	Besides the severe challenge of non-stationarity, it is in general difficult to deduce to what extent an agent contributed to state transitions and thus the rewards received as each agent is capable of manipulating the environment. This is known as the \textit{multi-agent credit assignment problem} (\cite{Chang.2004}). Additionally, the exploration-exploitation trade-off common to RL even worsens in the cooperative case. This is due to other learning agents which might concurrently explore. \cite{Matignon.2012} refer to this problem as \textit{alter-exploration}. Furthermore, the \textit{coordination problem} states that successful cooperation requires the agents to coordinate their controls in order to avoid e.g. shadowed equilibria (\cite{Matignon.2012}).
	Finally, in order to cope with the majority of control problems, we require compatibility with continuous state and control spaces and nonlinear systems and do not assume restrictions concerning the structure of the agents' objectives. 	
	However, as a system model is usually available in control engineering as a result of model design or an identification process, we desire to incorporate this beneficial knowledge into our method. Due to causality, we assume that the joint\tdd{ACHTUNG! Mit Notation der mathematischen Problemstellung abgleichen (aggregatesd control signal)} control signals of other agents are not instantaneously measurable at run time but are retrospectively measurable or deducible. \commd{Den folgenden Satz löschen, da es eigentlich doch nicht so resonable ist, wenn man wirklich die Stellgröße jedes einzelnen anderen Agenten braucht und nicht nur die ``Summenstellgröße" oder ähnliches? }
	
	\subsection{Related Work}\label{sec:related_work}
	In the following, a short overview regarding related work concerning cooperative control will be given and analyzed w.r.t. our problem. One possible approach as proposed by \cite{Kopf.2019b} is to identify and constantly update the aggregated control law of all other agents. Relying on a model of the system dynamics, this allows a simulation-based optimization of the cooperative control problem. The concept of opponent or partner modeling is also discussed by  
	\cite{Lowe.2017} (Section~4.2 therein). When facing a dynamic game setting, another approach to cope with unknown partners in cooperative scenarios is given by the identification of associated cost functionals as done by \cite{Kopf.2017} and \cite{Inga.2018} and a subsequent optimization. In the human-machine context, this setting is motivated by the assumption that human motion can be modeled by means of optimal control \citep{Scott.2004}.
	
	
	In contrast to these methods, the following approaches avoid the need to identify the partners' cost functionals or control laws. Among these methods, Adaptive Dynamic Programming in the Cooperative Control setting (\cite{Vamvoudakis.2011, Kopf.2018}) focuses on efficient adaptation from a control-oriented perspective but has more restricting assumptions regarding reward structures and system dynamics compared to deep RL methods. Thus, the following methods either rely on extensions to \textit{Deterministic Policy Gradient} (DPG) methods (\cite{Silver.2014b}) or extensions to \textit{Deep Q-Networks} (DQN) (\cite{Mnih.2015}). 
		
Among the DPG methods, either all agents need to know the policy parameters of all others \citep{Gupta.2017}, explicit opponent modeling is required when facing our problem (cf. \cite[Section~4.2]{Lowe.2017}), or all agents share the same critic and a global reward function \citep{Foerster.2018b}, i.e. the agents are not decentralized and fully cooperative. Furthermore, the DPG based methods suffer from increasing variance in multi-agent domains (cf. \cite{Lowe.2017} and \cite{Foerster.2018b}), which destabilizes the training process particularly with independently learning agents.		
	Concerning the DQN-based methods, they either work in the fully cooperative setting with finite state and control spaces (\cite{Foerster.2017c}; \cite{Matignon.2007}), are limited to finite control spaces \citep{Omidshafiei.2017} or finite state and control spaces \citep{Palmer.2018}.

	
	\subsection{Contributions of This Paper}\label{sec:contribution}
	As none of the deep MARL methods in literature fulfills our control-oriented requirements, we propose a new approach for cooperative control in continuous state and control spaces. Our method does not depend on the explicit identification of the other agents' behavior. Instead, an adapting automation is explored, which is not reliant on the premise of other agents behaving optimally and is expected to facilitate a high degree of generalizability across domains and partners. Compared to recent deep RL methods in the multi-agent domain, we face the challenge of decentralized agents with no knowledge of the partners' control strategies or objectives and no explicit means of communication and present three new, modular mechanisms which explicitly address the associated challenges.
	
	Although the deep MARL methods in Section~\ref{sec:related_work} cannot applied be directly to our problem setting, they reveal reoccurring mechanisms which we rely on: First, extensions to the \textit{experience replay memory} (ERM) in order to counteract the difficulty of applying experience replay in non-stationary environments. Second, \textit{variable learning rates} in order to induce coordination and facilitate the use of a temporal dimension in the sampling process.
	We propose \textit{Temporal Experience Replay} (TER) to account for the non-stationarity of the environment each agent faces. The main idea behind TER is a time-dependent prioritization of samples in the experience replay memory. 
	Furthermore, we introduce the idea of \textit{Imagined Experience Replay} (IER), which benefits from a model of the system dynamics and grounds the training process by means of fictional experiences. IER can be understood as an adaptation of the idea of imagination rollouts (cf.~\cite{Gu.2016}) to cope with the challenges encountered in multi-agent settings. In addition, in order to address the multi-agent credit assignment problem, we propose a new mechanism of variable learning rates. Our accountability-driven approach termed \textit{impact Q-learning} (IQL) ties the learning rate to the agent's contribution towards the joint control.	
	We further combine IQL and IER to simulate targeted cooperation scenarios in order to exhaust potential coordination between agents. Finally, the mechanisms are made dependent on an exploration rate such that the influence of each distinct concept is varied according to its current utility. This increases their effectiveness and reduces issues connected to alter-exploration.

	\commd{Either the introduction includes the state of the art (related work) OR this can be done in another section (before or after the Problem Definition, wherever suited best). $\rightarrow$ included in introduction as this facilitates the argumentation regarding novelty/necessary of our method.}

	\tdd{evtl. ``Structure of the Paper"; erst recht spät schreiben, SEHR kurz halten; vielleicht auch garnicht notwendig.}

	\section{Formal Problem Definition and Prerequisites}\label{sec:problem}
	We now formalize our problem definition and introduce prerequisites on which our proposed mechanisms rely on.
	\subsection{Formal Problem Definition}
	Consider a discrete-time system $f:X\times U\rightarrow X$ that is controlled by $N$ agents given by
	\begin{align}\label{eq:system}
	x_{k+1}=f(x_k, u_{1,k}, \dots, u_{N,k}),
	\end{align}
	where $x_k\in X\subseteq\mathbb{R}^n$ denotes the state at time step $k$, $u_{i,k}\in U_i\subseteq\mathbb{R}$ the control of agent $i\in\mathcal{N}=\left\{1,\dots,N\right\}$ and $U=U_1\times \dots \times U_N$ the joint control space. Depending on the current state $x_k$ and controls $u_{i,k}$, each agent $i\in\mathcal{N}$ experiences a reward $r_i$ that results from a reward function $g_i:X\times U\rightarrow \mathbb{R}$, i.e.
	\begin{align}
	r_{i,k} = g_i(x_k,u_{1,k},\dots,u_{N,k}).
	\end{align}
	The goal of each agent is to adapt his control law $\pi_i:X\rightarrow U_i$ in order to maximize his value
	\begin{align}\label{eq:return}
	V_i^{\m{\pi}}(x_k) = \sum_{k=0}^{\infty}\gamma_i^k r_{i,k}=\sum_{k=0}^{\infty}\gamma_i^k g_i(x_k, \pi_1(x_k),\dots,\pi_N(x_k)),
	\end{align}
	i.e. the long-term discounted reward under the tuple of control laws $\m{\pi}=\left(\pi_1,\dots,\pi_N\right)$, where $\gamma_i\in[0,1)$ denotes a discount factor. Thus, our deterministic game setting (in contrast to the stochastic game definition in \cite{Busoniu.2010e}) is defined by the tuple $G=\left(X, U_1, \dots, U_N, f, g_1, \dots, g_N, \gamma_1,\dots,\gamma_N\right)$. Our problem is then formalized as follows.
	\begin{prob}\label{prob:problem}
		Given the game $G$, each agent $i\in\mathcal{N}$ knows the system dynamics $f$ and his own reward function $g_i$. Furthermore, each agent $i\in\mathcal{N}$ receives his current reward $r_{i,k}$ at time step $k$ and is able to deduce the \textit{previous} controls $u_{j,k-1}$, $\forall j\in\mathcal{N}\setminus\{i\}$ of other agents but has no access to the current controls $u_{j,k}$, other agents' control laws $\pi_j$, their reward functions $g_j$ or actual rewards. In this setting, each agent $i\in\mathcal{N}$ aims at adapting his control law $\pi_i$ in order to maximize $V_i^{\m{\pi}}$ as defined in \eqref{eq:return}. 
	\end{prob}

	\subsection{Prerequisites Concerning Deep Q-Networks}
	Our algorithm is based on DQN. Thus, the fundamental concepts of Q-learning and DQN are introduced in the following. Q-learning (\cite{Watkins.1989}) is an iterative algorithm which intends to learn an optimal state-action-value function $Q^*$. Here,
	\begin{align}
	Q^*(x_k,u_k)=\max_{\pi}Q^{\pi}(x_k,u_k)
	\end{align}
	holds, where $Q^{\pi}(x_k,u_k)$ represents the discounted long-term cost, if an agent is in state $x_k$ and applies the control, i.e. action, $u_k$ at time step $k$ and follows the control law $\pi$ thereafter.  
	The relevance of $Q^*$ becomes clear as the optimal control law maximizing the long-term discounted reward (cf. \eqref{eq:return} for $N=1$) is given by
	\begin{align}\label{eq:pi_star}
	\pi^*(x_k)=\argmax_{u_k}Q^*(x_k,u_k).
	\end{align}
	The update rule in order to estimate $Q^*$ is given by
	\begin{align}\label{eq:q_update}
	\begin{aligned}
	Q(x_k,u_k)\leftarrow\, & Q(x_k,u_k)\\&+\alpha_k\underbrace{\left[r_k+\gamma\max_{u}Q(x_{k+1},u)-Q(x_k,u_k)\right]}_{\delta_k},
	\end{aligned}
	\end{align}
	where $\delta_k$ denotes the temporal difference (TD) error and $\alpha_k\in(0,1]$ a learning rate. The TD error $\delta_k$ thus measures the difference between the current Q-function estimate $Q(x_k,u_k)$ and the TD target $r_k+\gamma\max_{u}Q(x_{k+1},u)$. The tuple $\chi_k=(x_k,u_k,r_k,x_{k+1})$ is taken from interaction with the environment.
	
	In order to extend Q-learning to continuous state spaces, function approximators such as deep neural networks which parametrize the Q-function have been introduced. Here, the work of \cite{Mnih.2015} marked a breakthrough, as the introduction of \textit{Experience Replay} (ER) significantly improved training. The idea is to randomize training samples in order to remove correlation between observed state-transition sequences. Therefore, experience tuples $\chi_k$ are stored in an ER memory (ERM) $\mathcal{M}$ at each time step $k$. A Q-learning update is then performed by sampling (e.g. uniformly at random) from the ERM and minimizing the associated squared TD error $\delta_k$. In order to account for continuous control spaces, \cite{Gu.2016} introduced the concept of \textit{Normalized Advantage Functions} (NAF), allowing to deduce an analytical expression in order to solve \eqref{eq:pi_star}.

	\commd{So wirklich zufrieden bin ich mit Abschnitt 2 noch nicht, also falls du noch Ideen hast, bin ich da offen!} \fsamd{@Samuel: sollte am Ende noch Double DQN eingeführt werden? Wird darauf später verwiesen/wird das benötigt?}
	
	\section{Decentralized Cooperative Control Method}\label{sec:our_method}
	\tdd{Samuel; bestenfalls direkt parallel in ``Prerequisites" vermerken, was konkret dort auftauchen muss, um die ``Proposed Method" sauber zu verstehen. Ich bin auch stark für $\m{x}$ und $\m{u}$ anstatt $\m{s}$ und $\m{a}$, da in Regelungstechnik viel üblicher! Zudem: bestenfalls zu Beginn schreiben, dass das für $N$ Agenten geht, hier aber aus Ego-Perspektive eines Agenten alle anderen Agenten als ein Partner zusammengefasst. }\commd{ich tendiere zu American English, da etwas  verbreiteter in der Community, wir sollten es insbesondere aber einheitlich machen}


In order to gain control of the challenges associated with Problem~\ref{prob:problem}, we propose a time-dependent mechanism termed \emph{Temporal Experience Replay} (TER) to account for the non-stationary environment, include known system dynamics by means of \emph{Imagined Experience Replay} (IER) and use variable learning rates with the proposed \emph{Impact Q-Learning} (IQL) in order to induce coordination. As these mechanisms can be applied in a modular fashion, they are separately introduced and then combined in Section~\ref{sec:algorithm}.

	\subsection{Temporal Experience Replay (TER)}\label{sec:TER}
	The proposed method of \emph{Temporal Experience Replay} attempts to unify the idea of favoring more recent experiences with the concept of more probable sampling of experiences according to a prioritization factor. \commd{wieder ein bisschen Stand der Technik...}
	Analogue to \emph{Prioritized Experience Replay} \citep{Schaul.2016}, \commd{@Samuel: Ich stimme dir zu, dass das an der Stelle von der Formulierung noch nicht ganz passt. PER erzeugt ja keinen bias zu späteren samples, sondern nur zu ``wichtigeren". }we suggest to bias the sampling process. However, instead of utilizing the TD error for the prioritization, we propose to focus towards recent experiences by introducing a \emph{temporal prioritization} $\tau$, which is proportional to the time that has passed since collection $k_c$ of the state-transition:
	\begin{equation}\label{temporal_prior}
	\tau_{k_c}(k)=\exp\big(-\lvert k-k_c\rvert \ \big)+\xi_{\text{temp}},
	\end{equation}
	
	with the sampling probability $P_{k_c}(k)$ given by
	\begin{equation}\label{TER}
	P_{k_c}(k)=\frac{\tau_{k_c}(k)}{\sum_l\tau_l(k)}.
	\end{equation}
	In \eqref{temporal_prior} the optional offset $\xi_{\text{temp}}$ can be used to ensure that experiences are sampled with non-zero probability and the term $k$ denotes the current time step. Hence, to compute \eqref{temporal_prior} and \eqref{TER} at runtime, the experience tuple has to be extended by the respective current time step, producing the new tuple:
	\begin{equation}
	\chi^*_k=(x_k,u_k,r_{k},x_{k+1},k).
	\end{equation}\commd{oder hinten: $k_c = k$?}
	The underlying idea is that agents are more capable of adjusting to ever changing policies of other agents by experiencing recent state-transitions tuples 
	more often than old ones. However, the TER as described by \eqref{temporal_prior} and \eqref{TER} is impractical, as it leads to two major issues:
	Firstly, similar to approaches that restrict the memory size itself, the proposed temporal prioritization suffers from biasing the ERM too much towards recent experiences. This can lead to over-fitting of an agent's policy\commd{wird das nicht durch $\xi$ vermieden?}. Secondly, the temporal prioritization increases the computational complexity of the sampling process to a degree that is not feasible in practice. This is due to the computation of the temporal prioritization $\tau_{k_c}(k)$ itself, as it has to be updated for each experience tuple at every time step.
	
	To overcome both of these issues, a two step sampling process is proposed. Initially, a \emph{macro-batch} $\mathcal{B}$ of size $B$ is sampled uniformly at random from the complete experience replay buffer $\mathcal{M}$. Subsequently, a smaller \emph{mini-batch} $\mathcal{T}$ of size $t$, with $t<B$, is sampled from $\mathcal{B}$ utilizing the temporally prioritized probabilities given in \eqref{TER}. By dividing the sampling process into two manageable parts, both of the above mentioned problems are solved. The macro-batch $\mathcal{B}$ is only of size $B$, thus, the computational complexity of calculating the temporal priorities $\tau_{k_c}(k)$ is equally reduced to $B$. Additionally, the initial macro-batch is sampled uniformly at random, which reduces the risk of overemphasizing experiences related to recent episodes.
	
	In order to account for the varying exploration rate $\varepsilon_k$ of agents at different stages of the training process, we propose an additional exploration rate dependency of $B$ yielding a time-dependent macro-batch size $B_k$. TER attempts to induce adaptation to other agents’ policy changes. Therefore, it is most effective when the partners' policies start to converge and are less influenced by exploration noise. Thus, experiences should be sampled uniformly at random during early training (i.e. when $\varepsilon_k\approx 1$), which can be achieved by choosing $B_k$ close to the mini-batch size $t$, whereas during later training stages, i.e. once $\varepsilon_k\rightarrow 0$, $B_k$ should approach the final macro-batch size $B$. Consequently, we choose
	\begin{equation}\label{eq:Bk}
	B_k=(B-t)(1-\varepsilon_k)+t.
	\end{equation}

	\subsection{Imagined Experience Replay (IER)}\label{sec:IER}
	\commd{inklusive Abgrenzung zu \cite{Gu.2016}; ggf. mit Einleitung abgleichen}
	The above mentioned augmentations to ER attempt to either stabilize the training process in order to make agents less susceptible to changing environment dynamics or bias learning towards recent experiences to enable agents to adapt to changes in the dynamics. In any case it is acknowledged that the other agents’ behavior is indissociable from the dynamics of the environment, which is generally a reasonable presumption given independent and decentralized agents. However, due to the assumption that a system model is available, it becomes possible to ground the training process through simulated experiences in which the partners' controls are marginalized leading to stationary environment dynamics. This is the fundamental idea of our second proposed modification to the ER, which is termed \emph{Imagined Experience Replay} (IER).
	
	The concept of IER was inspired by the \emph{imagination roll-outs} developed by \cite{Gu.2016}, who proposed the idea of accelerating the training process by utilizing a learned model to simulate artificial experiences that were then added to the replay buffer. Differently, IER is used here to simulate experiences, which would not occur under normal circumstances. Specifically, all other agents' controls $\boldsymbol{u}_{-i}=\left\{u_1, \dots, u_{i-1}, u_{i+1}, \dots, u_N\right\}$ are set to zero while retaining the agent's control $u_i$ unchanged. Given a regular experience
	\begin{equation}
	\chi_{i,k}=(x_k,u_{i,k},r_{i,k},x_{k+1}) 
	\end{equation}
	for agent $i$ which occurred at time $k$, the successor state $x_{k+1}$ and received reward $r_{i,k}$ can be substituted by utilizing the underlying system dynamics $f$ and reward function ${g}_i$:\footnote{For convenience of notation, ${g}_i(x_k,u_{i,k},\boldsymbol{u}_{-i,k})$ and ${p}(x_k,u_{i,k},\boldsymbol{u}_{-i,k})$ evaluate $g_i(\cdot)$ and $p(\cdot)$ at the state $x_k$ while agent $i$ applies the control $u_{i,k}$ and all other agents controls are denoted by the tuple $\boldsymbol{u}_{-i,k}$.}
	\begin{equation}
	\chi_{i,k}=\big(x_k,u_{i,k},\ {g}_i(x_k,u_{i,k},\boldsymbol{u}_{-i,k}),\ {p}(x_k,u_{i,k},\boldsymbol{u}_{-i,k})\big).
	\end{equation}
	Subsequently, an \emph{imagined experience} $\tilde{\chi}$ can be simulated by replacing the other agents' controls $\boldsymbol{u}_{-i}$ by $\boldsymbol{0}$ yielding the imagined successor state $\tilde{x}_{k+1}$ and reward $\tilde{r}_{i,k}$:
	\begin{equation}\label{IER}
	\tilde{\chi}_{i,k}=\big(x_k,u_{i,k},\ \underbrace{{g}_i(x_k,u_{i,k},\boldsymbol{0})}_{\tilde{r}_{i,k}},\ \underbrace{{p}(x_k,u_{i,k},\boldsymbol{0})}_{\tilde{x}_{k+1}}\big).
	\end{equation}
	In contrast to the imagination roll-out of \cite{Gu.2016}, the imagined experiences in \eqref{IER} are not stored to the actual ERM and sampled from there. Instead, an exploration rate dependent probability $\tilde{P}$ is utilized to determine, whether an imagined experience is computed in addition to the sampled, observed experience. Once used for training, the imagined experience is discarded in order to reduce the risk of overemphasizing artificial experiences in which partners are non-existent.
	\commd{ist es hier sinnvoll eine Unterteilung in 3.2.1 und 3.2.2 vorzunehmen? Denke eher nicht, wäre zwar übersichtlich, aber noch platzraubender}	
	During the initial training phase, agents predominantly explore random controls. 
	Hence, it is not possible to infer other agents’ policies from observations, and the application of IER, in order to stabilize the training process, is most useful, as experiences are simulated in which solely the agent $i$ interacts with the environment. These imagined experiences, at this stage of training, are essentially observations for which the exploration noise of other agents is not present. Consequently, the probability $\tilde{P}$ of simulating imagined experiences $\tilde{\chi}_{i,k}$ at time step $k$ is proposed to be proportional to the current exploration rate $\varepsilon_k$, i.e. $\tilde{P}(k)\sim\varepsilon_k$.
	
	However, during the later stages of training, when policies start to converge
	and are less influenced by exploratory controls, the coordination between agents and the adaptation to the partners' policies becomes more important. Upon closer examination it can be seen that by generating artificial experiences in which cooperation between agents is simulated, the IER can potentially be utilized to induce coordination between agents. Opposite to \eqref{IER}, these \emph{imagined coordination experiences} are more useful during later stages of training. Therefore, the respective sampling probability is proposed as $P_{\text{coord}}(k)\sim(1-\varepsilon_k).$
	When generating experiences with the purpose of inducing coordination, it has to be considered that the final algorithm is required to entail mixed cooperative-competitive task types. Thus, it cannot generally be presumed that the agents’ respective goals are compatible. Consequently, it is proposed that the IER is utilized to simulate three additional scenarios:
	\begin{enumerate}
		\setcounter{enumi}{0}
		\item In order to induce coordination, the control $u_i$ of agent~$i$ is discarded, causing it to be idle:
		\begin{align}\label{coop_0}
		\chi^{(i)}_{\text{idle}} &=\big(x,0,\ {g}_i(x,0,\boldsymbol{u}_{-i}),\ {p}(x,0,\boldsymbol{u}_{-i})\big).
		\end{align}
		Therefore, by utilizing $\chi^{(i)}_{\text{idle}}$, agent $i$ can observe how other agents behave by themselves and whether the resulting environment transitions are beneficial.
		\item Here, the first of two cooperation scenarios $\chi^{(i)}_{\text{coop1}}$ is simulated. For this purpose, the agent's controls $u_i$ are set to be equal to the average of all other agents' joint control $\overline{{u}\raisebox{2mm}{}}_{-i}$:
		\begin{equation*}
		\overline{{u}\raisebox{2mm}{}}_{-i}=\frac{1}{N-1}\sum_{\forall j\in\mathcal{N}\setminus\{i\}}^{}u_j,
		\end{equation*}
		resulting in:
		\begin{align}\label{coop_1}
		\chi^{(i)}_{\text{coop1}} &=\big(x,\overline{{u}\raisebox{2mm}{}}_{-i},\  {g}_i(x,\overline{{u}\raisebox{2mm}{}}_{-i},\boldsymbol{u}_{-i}),\  {p}(x,\overline{{u}\raisebox{2mm}{}}_{-i},\boldsymbol{u}_{-i})\big).
		\end{align}
		\item In the second cooperation scenario $\chi^{(i)}_{\text{coop2}}$, the inverse is generated. Each element of the other agents' joint control $\boldsymbol{u}_{-i}$ is modified to be equal to $u_i$:
		\begin{align}\label{coop_2}
		\chi^{(i)}_{\text{coop2}} &=\big(x,u_{i},\ \!  {g}_i(x,u_{i},\boldsymbol{u}_{-i}),\ \! {p}(x,u_{i},\boldsymbol{u}_{-i})\big), \\ \nonumber &\qquad \text{with}\ u_{j}\!\coloneqq u_{i}, \forall j\in\{1,\ldots,N\}.
		\end{align}
	\end{enumerate}
	
	By imagining these three scenarios, potential coordination possibilities are exhausted. The first one specifically enables agents to evaluate whether being idle leads to an acceptable reward, which in competitive environments is generally discouraged and will be evaluated correspondingly, whereas the second and third evaluate the effect if agent $i$ imitates the average control of the others or if all agents stick to the control of agent $i$.
	
	\subsection{Impact Q-Learning (IQL)}\label{sec:IQL}
	
	\commd{Wo ``targeted cooperation" (Zusammenspiel IER und IQL) einbringen?}
	In the previous subsections, different additions to the ERM were proposed, which focused mainly on providing stability to counteract the problem of a non-stationary environment. An additional, often utilized mechanism in MARL are variable learning rates. The hysteretic (\cite{Matignon.2007}) and lenient (\cite{Palmer.2018}) method  are two representatives of optimistic learners, which are generally well suited to induce coordination. However, the core principle of optimistic agents, which reduce their learning rate given negative experiences, is diametrically opposed to the challenge of multi-agent credit assignment. In order to credit agents correctly with respect to the observed outcome, it is necessary for them to not only learn notably from positive experiences but also from negative ones. Furthermore, the concept of tying the learning rate to rewards (\cite{Matignon.2007}) is itself flawed to combat credit assignment, as the attention an agent should pay to certain experiences ideally does not depend on the quality of the outcome, but on the contribution of an agent towards the observed outcome.\commd{wieder sehr langer Abschnitt, evtl. auch in neu strukturierten Herausforderungen/Stand der Technik teil integrieren? sehr knifflig!}
	
	Therefore, we propose a novel approach for solving the multi-agent credit assignment problem using variable learning rates. We attempt to tie the variable learning rate to the actual contribution of an agent towards the observed state transitions. This is facilitated by the retrospective observation of all agents' controls, as it enables each agent to compare its control $u_{i,k}$ at time step $k$ to the separately remaining joint control $\boldsymbol{u}_{-i,k}$ of all other agents. To this end we introduce a novel quantity, called \emph{impact factor}
	\begin{equation}\label{impact_factor}
	\lambda_{i,k}=\frac{\lvert u_{i,k} \rvert}{\sum\limits_{j=1}^{N} \lvert u_{j,k} \rvert},
	\end{equation}
	which describes the agent's relative contribution to the joint control, and thus, to experienced state transitions.
	In order to enable the computation of a meaningful impact factor $\lambda$ in \eqref{impact_factor} it is presupposed for IQL that agents share the same control space $U_1=\ldots=U_N$ and that all agents' controls manipulate the system equally. Subsequently, the update rule \eqref{eq:q_update} for an agent $i$ can be modified to apply different learning rates depending on the agent's impact factor:
	\begin{align}\label{impact_update}
	\delta_k&\leftarrow r_{k}+\gamma\max_{u}Q_{i}(x_{k+1},u)-Q_{i}(x_{k},u_k), \nonumber \\ 
	Q_{i}(x_{k},u_k)&\leftarrow
	\begin{cases}
	Q_{i}(x_{k},u_k)+\alpha\delta_k & \text{if}\ 1.0\geq\lambda_{i,k}>\lambda_{\text{high}}\\
	Q_{i}(x_{k},u_k)+\sigma\delta_k & \text{if}\ \lambda_{\text{high}}\geq\lambda_{i,k}\geq \lambda_{\text{low}}\\
	Q_{i}(x_{k},u_k)+\beta\delta_k & \text{if}\ \lambda_{\text{low}}>\lambda_{i,k}\geq 0,
	\end{cases}
	\end{align}
	with $0<\beta<\sigma<\alpha<1$.
	\commd{Möglicherweise ist es sauberer die Zahlen 0.8 und 0.2 nicht zu verwenden und stattdessen von generischen $\lambda_{\text{high}}$ und $\lambda_{\text{low}}$ zu sprechen. Finde ich gut, dann muss aber im Ergebnisteil noch erwähnt werden, wie beide Größen gewählt wurden.}
	In \eqref{impact_update}, the Q-learning update rule is partitioned into three distinct impact ranges with which it is possible to differentiate whether an agent had a high, medium, or low influence towards a state transition.
	Hence, the amount an agent learns from an experience is proportional to its respective contribution or impact.
	When a positive experience is observed, the Q-value estimate is only increased heavily, if the agent can be credited for the event. On the other hand, when a punishment occurs, the agent is mainly discouraged from the corresponding state-action pair, if the agent is at least in part responsible. Particularly this kind of accountability-driven learning behavior is required for agents to overcome the credit assignment challenge.
	
	\subsection{Algorithm}\label{sec:algorithm}
	Upon closer examination of IER, it can be seen that the concept of simulated experiences specifically for the coordination scenarios in \eqref{coop_0}, \eqref{coop_1}, and \eqref{coop_2} may reduce the coordination problem’s severity, but also produces additional computational effort. Thus, it is advisable to limit these calculations to state-action pairs with high potential for coordination. This can be done by utilizing the computed impact factors. In \eqref{impact_update}, three intervals with different degrees of an agent's impact were distinguished. When analyzing the ones corresponding to learning rates $\alpha$ and $\beta$, it can be seen that the potential for coordination is limited here, because the agent either predominantly contributes towards the state-transition or only has a minor impact. However, for the case of medium learning rates $\sigma$, the impact of agents, particularly in the case of only few agents\commd{Ignorieren wir die spezifische Eignung für den 2-player case insgesamt? Ggf. weiter oben schon erwähnen, dass der Fokus auf "few agents" liegt?}, is distributed more evenly, which in turn increases the need for coordination. In this instance, the simulation of different IER scenarios is most powerful and the trade-off between computational effort and induced coordination most beneficial.
	Further, two distinct kinds of medium-impact experiences are distinguished. Either the agent's control $u_{i,k}$ and the average of all remaining controls $\overline{u\raisebox{2mm}{}}_{-i,k}$ work in the same direction, or against each other. This can be determined by sampling an experience $\chi_k$ and computing a \emph{coordination coefficient} $\psi_k$ as such:
	\begin{equation}\label{coordination coefficient}
	\psi_{i,k}=\text{sgn}(\overline{u\raisebox{2mm}{}}_{-i,k}\cdot u_{i,k}).
	\end{equation}
	If $\psi_{i,k}$ equals 1, agent~$i$ and the others work in the same direction and it is not necessary to simulate the coordination experiences $\chi_{\text{idle}}$, $\chi_{\text{coop1}}$ and $\chi_{\text{coop2}}$. Instead, the learning rate $\sigma$, which is normally used for $\lambda_{\text{high}}\geq\lambda_{i,k}\geq \lambda_{\text{low}}$\commd{hier dann auch $\lambda_{\text{high}}$ bzw. $\lambda_{\text{low}}$ verwenden} in \eqref{impact_update}, is substituted by the larger learning rate $\alpha$. Therefore, agents are induced to emphasize cooperative experiences during the learning process. In the case that agents act in opposing directions, $\psi_{i,k}$ equals -1. Besides the sampled experience $\chi_k$, the artificial experiences $\chi_{\text{idle}}$, $\chi_{\text{coop1}}$ and $\chi_{\text{coop2}}$ are simulated, and subsequently, the agent is trained on all of them. Here the lowest learning rate $\beta$ is applied, because the trained on experiences have not actually occurred and are only imagined for coordination purposes.	
	Thus, the instances for which the computational strenuous task of simulating multiple coordination experiences is required, can be reduced greatly and focused to occasions connected to the highest expected learning progress. The resulting algorithm after finally assembling the above described mechanisms is described in Algorithm~\ref{algo1}, where $U(a,b)$ denotes a uniform distribution in the intervall $[a,b]$.
	\algnewcommand\algorithmicforeach{\textbf{for each}}
	\algdef{S}[FOR]{ForEach}[1]{\algorithmicforeach\ #1\ \algorithmicdo}
	\begin{algorithm}[tb!]
		\caption{Deep impact Q-learning with TER and IER}\label{algo1}
		\begin{algorithmic}[1]
			\State \textbf{Input: }macro-batch size $B$, mini-batch size $t$,
			\State \hphantom{\textbf{Input: }}learning rates $\alpha$, $\sigma$, and $\beta$, ERM size $M$,
			\State \hphantom{\textbf{Input: }}target update frequency $m$, decay rate $\varpi$, 
			\State \hphantom{\textbf{Input: }}minimum exploration rate $\varepsilon_{\text{min}}$, maximum
			\State \hphantom{\textbf{Input: }}number of episodes $E_\text{max}$ and maximum
			\State \hphantom{\textbf{Input: }}time steps per episode $K_{\text{max}}$
			\State \textbf{Initialize: }$Q(x,u;\theta)$ and $Q(x,u;\hat{\theta})$ with random  
			\State \hphantom{\textbf{Initialize: }}weights $\theta$ and $\hat{\theta}$, ER buffer $\mathcal{M}\gets\varnothing$ with 
			\State \hphantom{\textbf{Initialize: }}size $M$, exploration rate $\varepsilon=1$
			\For {episode $e=1,\dots,E_\text{max}$}
			
			\State $k=0$
			\While{episode not terminated and $k\leq K_{\text{max}}$}

			\State With probability $\varepsilon$ select \textbf{random control} $u_k$
			\State Otherwise select $u_k=\argmax_{u}Q(x_k,u;\theta)$
			\State Execute $u_k$ and observe $r_{k}$, $x_{k+1}$
			\State Store tuple $(x_k,u_k,r_{k},x_{k+1},k)$ in $\mathcal{M}$
			\State Compute $B_k=(B-t)(1-\varepsilon_k)+t$ \hfill\eqref{eq:Bk}
			\State Sample uniformly at random $\mathcal{B}$ of size $B_k$ 
			\State Compute $\tau_{k_c}(k)$ for transition in $\mathcal{B}$ \hfill\eqref{temporal_prior}
			\State Sample $\mathcal{T}$ of size $t$ $\sim$ $P_{k_c}(k)=\tau_{k_c}(k)/\sum_l\tau_l$ \hfill\eqref{TER}
			\ForEach {$\chi \in \mathcal{T} $}
			\State Extract time of collection $c$ from $\chi$
			\State Draw random variable $w\sim U(0,1)$
			\If {$w < \varepsilon_c$ (exploration rate at time $c$)}
			\State Compute $\tilde{\chi}_{c}=\big(x_c,u_{c},\ \tilde{r}_{c},\ \tilde{x}_{c+1}\big)$ 
			\State Set $\tilde{y}_c=\tilde{r}_{c}+\gamma\max_{u}Q(\tilde{x}_{c+1},u;\hat{\theta}\ \! )$
			\State Update $\theta$ with learning rate $\beta$ for  $\tilde{y}_c$
			\EndIf
			\State Compute $\lambda_{c}=\lvert u_{i,c} \rvert / \sum_j \lvert u_{j,c} \rvert$\hfill\eqref{impact_factor}
			\State Set ${y}_c={r}_{c}+\gamma\max_{u}Q({x}_{c+1},u;\hat{\theta}\ \! )$
			\If {$\lambda_{c} > \lambda_{\text{high}}$}\commd{hier dann auch $\lambda_{\text{high}}$ bzw. $\lambda_{\text{low}}$ verwenden}
			\State Update $\theta$ with learning rate $\alpha$ for ${y}_c$
			\ElsIf {$\lambda_{\text{high}} \geq \lambda_{c} \leq \lambda_{\text{low}}$}\commd{hier dann auch $\lambda_{\text{high}}$ bzw. $\lambda_{\text{low}}$ verwenden}			
			\If {$\sgn{(\overline{u\raisebox{2mm}{}}_{-i,c}\cdot u_{i,c})}\geq 0$}
			\State Update $\theta$ with learning rate $\alpha$ for ${y}_c$
			\Else
			\State Update $\theta$ with learning rate $\sigma$ for ${y}_c$
			\EndIf
			\If {$\sgn(\overline{u\raisebox{2mm}{}}_{-i,c}\cdot u_{i,c})< 0$ \textbf{and} $\varepsilon_c < w$}
			\State Compute $\chi_{\text{idle},c}$, $\chi_{\text{coop1},c}$, $\chi_{\text{coop2},c}$
			\State Set target ${y}_{\text{idle},c}$, $y_{\text{coop1},c}$, and $y_{\text{coop2},c}$
			\State Update $\theta$ with learning rate $\beta$ for \State ${y}_{\text{idle},c}$, $y_{\text{coop1},c}$, and $y_{\text{coop2},c}$
			\EndIf
			\Else
			\State Update $\theta$ with learning rate $\beta$ for ${y}_c$
			\EndIf
			\EndFor
			\State Every $m$ steps, update target network: $\hat{\theta}\leftarrow\theta$
			
			\State $k=k+1$
			\EndWhile
			
			\State Decay exploration rate: $\varepsilon\leftarrow \text{max}[\varpi\cdot\varepsilon;\ \varepsilon_{\text{min}}]$\commd{wo $\varepsilon_{\text{min}}$ definiert?}
			\EndFor
		\end{algorithmic}
	\end{algorithm}
	
	\commd{Inklusive Abhängigkeiten von der exploration rate??}
	
	\tdd{Notation vereinheitlichen! Vorschlag der am häufigsten auftretenden Größen:\\$n$: Dimension Systemzustand. \\$N$: Anzahl Agenten. \\$x$: Zustand. \\$X$: Zustandsraum. \\$u_i$: Stellgröße Spieler $i$. \\$U_i$: Aktionsraum Spieler $i$. \\$g_i$: Belohnungs-/Kostenfunktion Spieler $i$. \\$r_i$: tatsächlicher Reward Spieler $i$. \\$\chi_{i}$: experience Spieler $i$. \\$f$: system dynamics. }
	
	\section{Results}\label{sec:results}
	In this section, the previously described algorithm is trained on a control task. Subsequently, the method's effectiveness is evaluated. 
	
	\commd{Simulation example, results, discussion}
	\subsection{Example System and Network Architecture}\label{sec:system}
	For the simulated environment, we use a customized two-player-variation of the \emph{OpenAI gym} (\cite{Brockman.2016}) cart-pole problem. Here, two agents balance a pole, which is hinged to a movable cart, by concurrently applying forces to the cart's base. The system dynamics are defined by the nonlinear differential equations
	\NewDocumentCommand{\qfrac}{smm}{%
		\dfrac{\IfBooleanT{#1}{\vphantom{\big|}}#2}{\mathstrut #3}%
	}
	\begin{align}
	\ddot{\theta}_k&=\frac{g\sin(\theta_k)-\cos(\theta_k)\Bigg[\qfrac{-F_{k, \text{res}}-m_{\text{pole}}\ \! l\ \! \dot{\theta}^{2}_k\ \! \sin(\theta_k)}{m_{\text{pole}}+m_{\text{cart}}} \Bigg]}{l\Bigg[\cfrac{4}{3}-\cfrac{m_{\text{pole}}\cos^2(\theta_k)}{m_{ \text{pole}}+m_{\text{cart}}}
		\Bigg]},\label{acc1}\\ 
	\ddot{s}_k&=\frac{F_{k,\text{res}}+m_{\text{pole}}\ \! l\ \! \big[\dot{\theta}^{2}_k \sin(\theta_{k})-\ddot{\theta}_k\cos(\theta_k) \big] }{m_{\text{pole}}+m_{\text{cart}}},\label{acc2}
	\end{align}
	where $g = \SI{-9.8}{\meter/\second^2}$, $m_{\text{pole}}=\SI{0.1}{\kilogram}$, $m_{\text{cart}}=\SI{1.0}{\kilogram}$, $l = \SI{0.5}{\meter}$ (half-pole length) and $F_{k,\text{res}}\in[\SI{-10}{\newton},\SI{10}{\newton}]$ (clipped sum of forces).
In \eqref{acc1} and \eqref{acc2}, $\theta_k$ denotes the angular displacement of the pole from \SI{0}{\radian}, which is defined by the pole standing perfectly upright. The cart's position is defined by $s_k$ with the center being at \SI{0}{\meter} and the system state is given by $x_k=\mat{s_k&\dot{s}_k&\theta_k&\dot{\theta}_k}^\intercal$. The successor state $x_{k+1}$ according to \eqref{eq:system} is calculated using the semi-implicit Euler method with a discrete time step of $\SI{0.02}{\second}$.

\begin{figure*}[t!]
	\centering
	\begin{subfigure}[t]{0.235\textwidth}\centering\includegraphics[width=\textwidth]{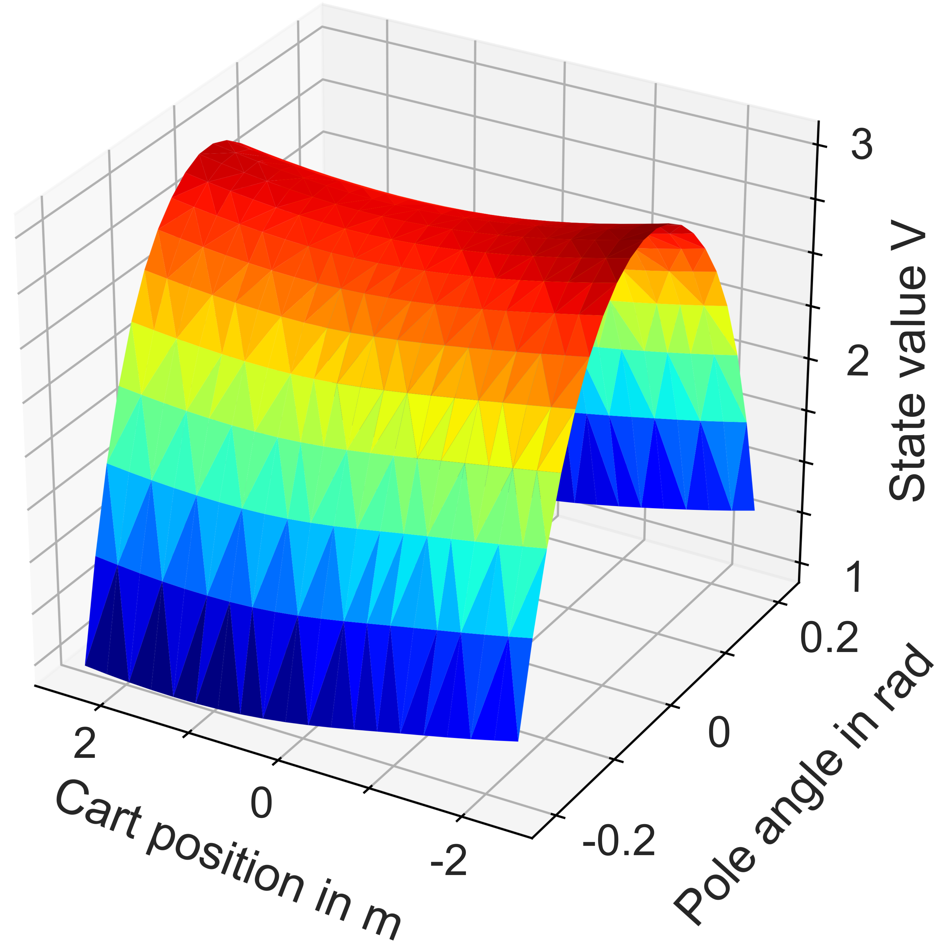}\caption{agent 1, 1000 episodes}\label{fig:1000ep_1}
	\end{subfigure}
	\begin{subfigure}[t]{0.235\textwidth}\centering\includegraphics[width=\textwidth]{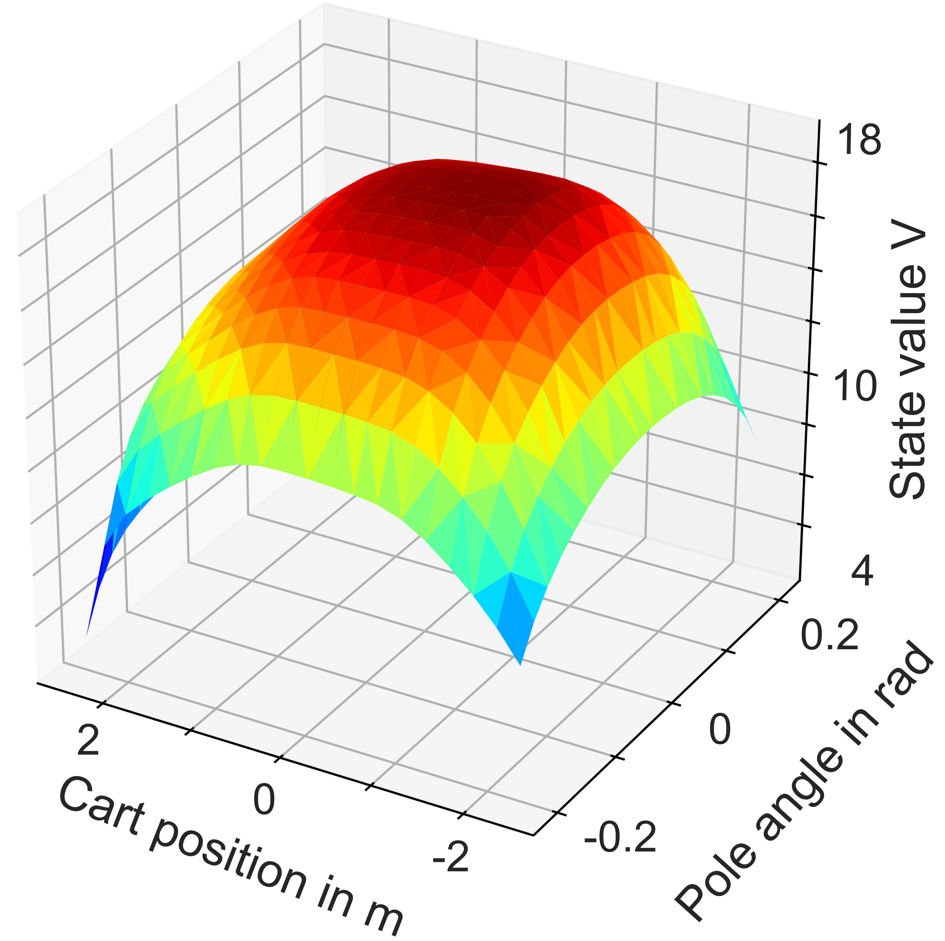}\caption{agent 1, 2000 episodes}\label{fig:2000ep_1}
	\end{subfigure}
	\begin{subfigure}[t]{0.235\textwidth}\centering\includegraphics[width=\textwidth]{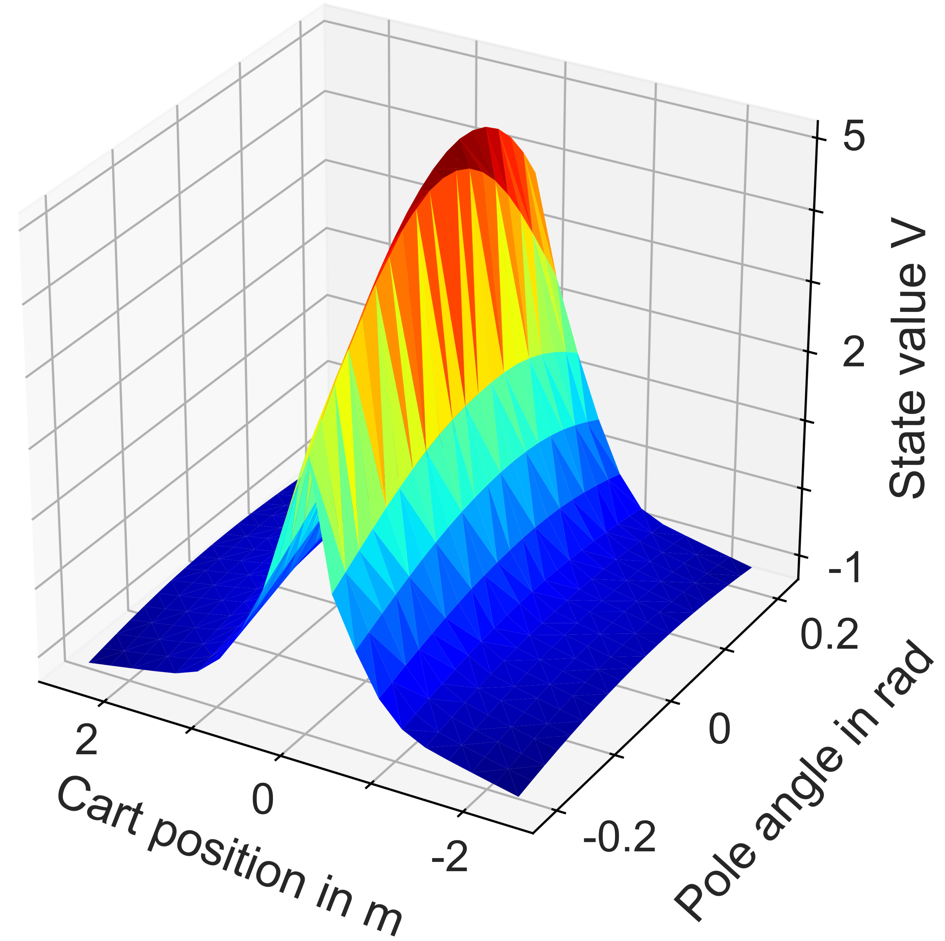}\caption{agent 2, 1000 episodes}\label{fig:1000ep_2}
	\end{subfigure}
	\begin{subfigure}[t]{0.235\textwidth}\centering\includegraphics[width=\textwidth]{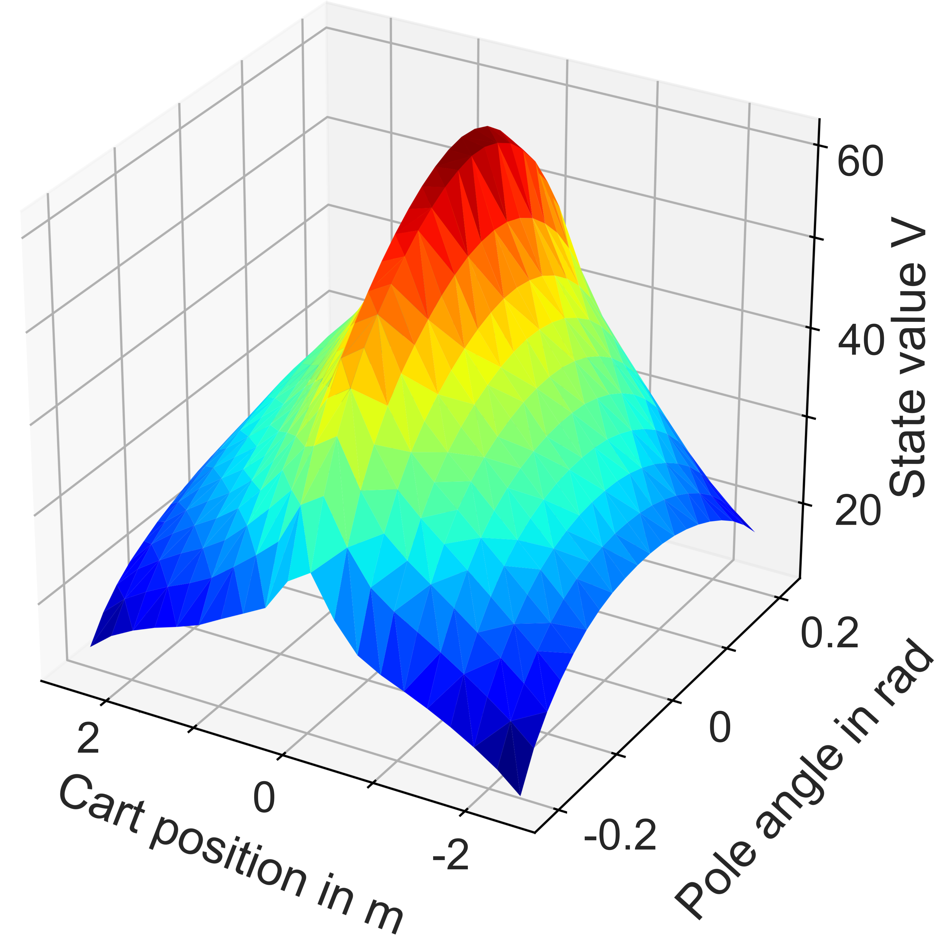}\caption{agent 2, 2000 episodes}\label{fig:2000ep_2}
	\end{subfigure}
	\caption{State-value estimates of the agents at different training stages. Each data point is averaged over $\dot{s}_k$ and $\dot{\theta}_k$.}
	\label{fig:V_agents}
\end{figure*}
	
%
%
%
	
	We assume that one agent focuses on balancing the pole upright, while the other agent is rewarded depending on the position of the cart. Thus, the first agent receives a reward $r_{1,k}$ of 1 for each time step $k$ in which the pole angle $\theta_k\in(\SI{-0.21}{\radian}, \SI{0.21}{\radian})$. If the episode is terminated, a reward of -1 is observed. On the contrary, the second agent's reward $r_{2,k}$ solely depends on the current cart position $s_k$. Specifically, a step-wise reward function is defined as such:
	\begin{equation}\label{reward_structure2}
	r_{2,k}=
	\begin{cases}
	+5, & \text{if}\ \lvert s_k-s^*\rvert< 0.1\ \! \text{m} \\
	+1, & \text{if}\ 0.1\ \! \text{m}\leq \lvert s_k-s^*\rvert<0.5\ \! \text{m} \\
	0, & \text{if}\ 0.5\ \! \text{m}\leq \lvert s_k-s^*\rvert<2.4\ \! \text{m} \\
	-1, & \text{if episode is terminated},
	\end{cases}
	\end{equation}
	with the target position denoted by $s^*$. We choose ${s^*=\SI{0}{\meter}}$. Agent~2 receives the highest reward in a small range around the target position, while the received reward is reduced step-wise once a certain boundary distance is exceeded.
	At the beginning of each of the $E_{\text{max}}=2000$ training episodes, the cart is initiated uniformly at random with the initial position $s_0\!\sim\! U(-\SI{2.3}{\meter},\SI{2.3}{\meter})$ and the initial pole angle $\theta_0\!\sim\! U(\SI{-0.085}{\radian},\SI{0.085}{\radian})$. An episode is terminated once one of the intervals $s_k\in\left[-2.4,2.4\right]\ \!$m or $\theta_k\in\left[-0.21,0.21\right]\ \!$rad is exceeded or $K_{\text{max}}=3000$ time steps have passed. We set $\gamma=0.999$.
	
	In our work, a dueling network architecture with NAFs as introduced by \cite{Gu.2016} was used. Additionally, multiple fully connected layers and dropout layers are stacked in front of the dueling network architecture to process observations. A description of the parameters corresponding to the network architecture and the hyperparameters used for training is given in the Appendix~\ref{sec:append_ANN_training_parameter}.
	
	\subsection{Simulations}
	\commd{3D plots value functions}
		\begin{figure}[t!]
		\centering
		\begin{subfigure}[t]{0.475\columnwidth}
			\centering
			\includegraphics[width=\columnwidth]{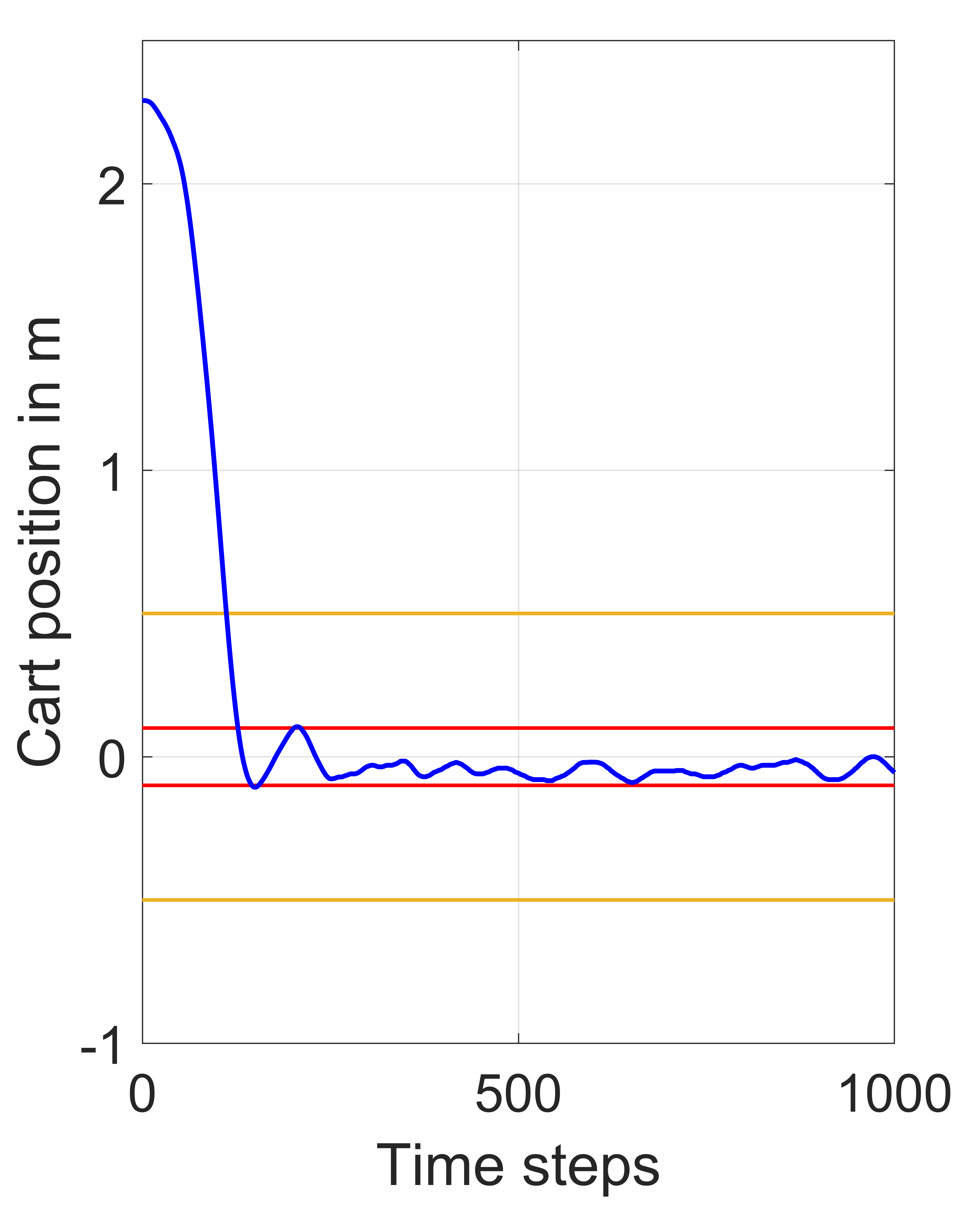}
			\caption{}
			\label{fig:PosControl1}
		\end{subfigure}
		\begin{subfigure}[t]{0.49\columnwidth}
			\centering
			\includegraphics[width=\columnwidth]{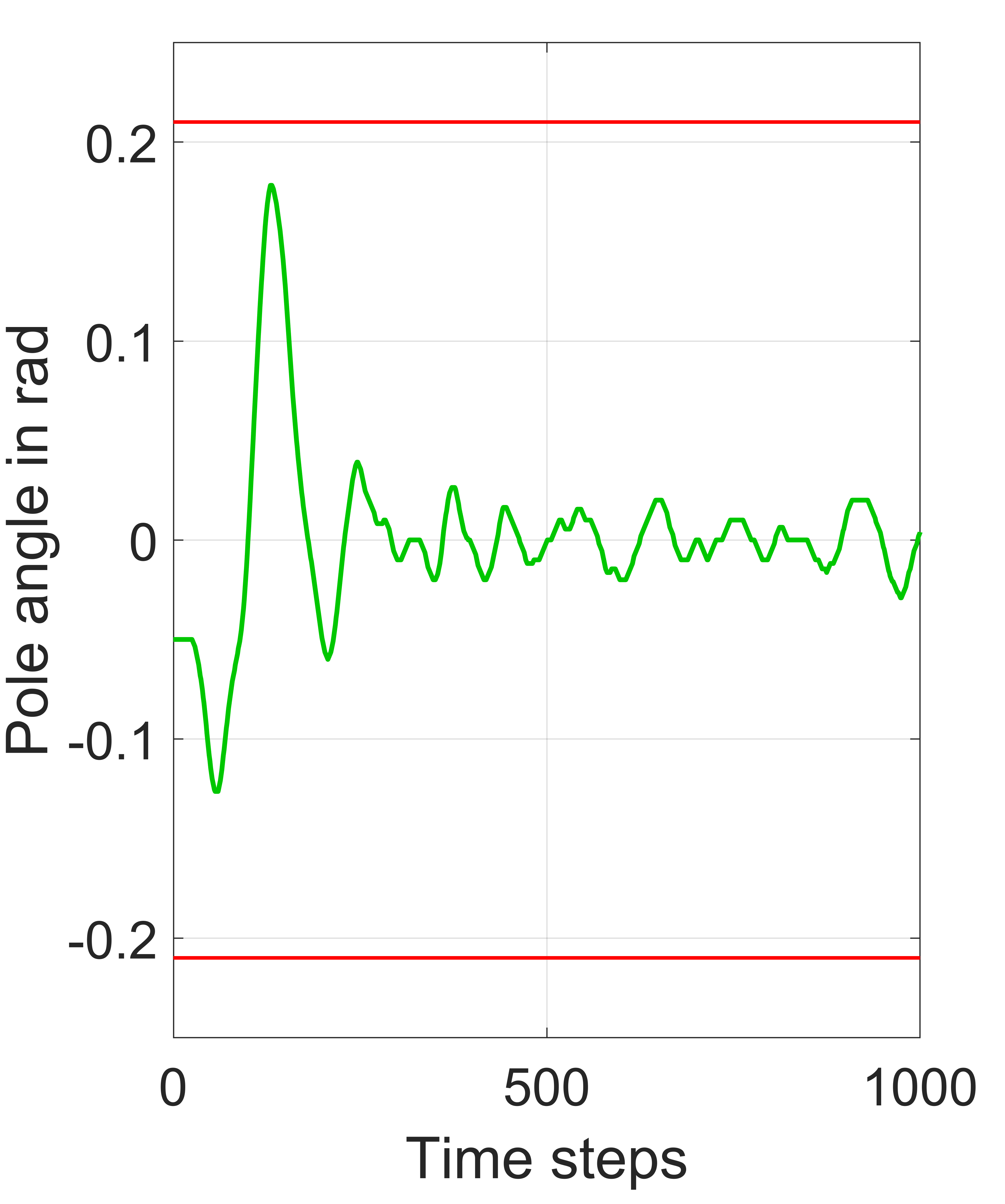}
			\caption{}
			\label{fig:PosControl2}
		\end{subfigure}
		\caption{Cart-pole position and pole angle for an example initialization using the learned controllers.}
		\label{fig:PosControl}
	\end{figure}
	

	Fig.~\ref{fig:V_agents} shows the resulting state-value estimates $V(x)$ of both agents at different training episodes, where each data point is averaged over $\dot{s}_k$ and $\dot{\theta}_k$ for reasons of presentability.
	After 1000 episodes of training, agent~1 expects the highest return along the pole angle of \SI{0}{\radian}. The lowest state-values are estimated for $\theta$ close to the terminal pole angles $\theta =\pm$\SI{0.21}{\radian}. Analogously, after 1000 episodes, agent~2 evaluates states close to the desired target position $s^*=\SI{0}{\meter}$ as most beneficial. This maximum state-value drops abruptly with slight deviation from $s^*$.
	Fig.~\ref{fig:2000ep_1} and \ref{fig:2000ep_2} show the agents' estimated state-values after 2000 episodes of training. For both agents, the expected returns have generally increased compared to the previous training stage. Because they successfully learned how to jointly balance the pole without moving the cart outside the boundaries, the available time steps to accumulate rewards is increased. Additionally, both agents learned to appropriately reduce $V(x)$ close to all terminal states.

	Example trajectories of the cart position $s_k$ and pole angle $\theta_k$ resulting from the trained control law are depicted in Fig.~\ref{fig:PosControl}. In Fig.~\ref{fig:PosControl1}, the boundaries for the highest 5 point reward range and the lower 1 point reward range for the seconds agent's position control are depicted in red and orange and the red lines in Fig.~\ref{fig:PosControl2} mark the terminal conditions.
	It can be seen that the agents are capable of moving the cart from the initial position to the desired target while holding the pendulum upright.
	
	\commd{comparison with DQN+NAF (no TER, IER, IF)?}

	\subsection{Discussion}
	After 1000 training episodes, the state-value estimations are predominantly dependent on the state dimension associated with the individual preferences as this yields the highest rewards while the cart-pole cannot be successfully controlled yet.	However, in later training stages (2000 episodes), the agents develop understanding concerning coordination possibilities (guided by IER) and their relative contribution (thanks to IQL) allowing the agents to move the cart to a desired state without terminating the episode yielding much higher rewards. \commd{optional}The decreased steepness of the state-value gradients when comparing $V(x)$ after 1000 and 2000 episodes is a result of the agent's increased control capabilities allowing to transition from a state with low rewards to a state associated with higher rewards.
	
	It is noticeable that without the mechanisms proposed in Section~\ref{sec:our_method}, the agents were not able to learn to stabilize the cart-pole at all with the given parametrization.
	Thanks to IER and IQL and an appropriate focus on recent experiences due to TER, the agents were successful at adapting to each other. The agents also learned to perform complex trajectories, which included deflecting the pole close to the terminal positions and angles (cf. Fig.~\ref{fig:PosControl}). Thus, it is possible for them to flexibly control the cart-pole even in difficult situations.

	\section{Conclusion}\label{sec:conclusion}
	In this paper, new mechanisms have been proposed in order to account for challenges arising in deep Multi-Agent Reinforcement Learning problems with restricted information. Two novel extensions to experience replay were presented. First, TER allows the sampling process to properly reflect the fact that recent experiences carry more information regarding the current control laws of cooperation partners and are thus better suited to counteract the non-stationarity compared to outdated experiences. Second, artificial experiences denoted as IER complement the experience replay memory. In the early training stage, alter-exploration problems are reduced due to simulated transitions in which the agents interact separately with the environment. Later, coordination is induced to exhaust the cooperation potential between agents as adaptation becomes feasible. Finally, these experience replay enhancements are supplemented by a mechanism termed IQL. Here, the relative contribution of the agent towards the observed outcome is accounted for by means of an impact factor which adapts an agent's learning rate.
	Our algorithm was evaluated on a simulated cart-pole-problem, where two agents successfully learned to cooperate.
	
	\commd{``Although a conclusion may review the main points of the paper, do not replicate the abstract as the conclusion. A conclusion might elaborate on the importance of the work or suggest applications and extensions."}

	
	\bibliography{mybib}             
	
	\appendix
	\section{Network Architecture and Training Parameter}\label{sec:append_ANN_training_parameter}    
	\commd{Die Hyperparameter müssen wir wohl nennen, auch wenn das leider ziemlich viel Platz benötigt. Abgespeckte Variante aus Anhängen~A.1 und A.2.}
	\begin{table}[h!]\captionsetup{width=\columnwidth}\caption{Hyperparameters of the network architecture}\label{tab:RL}
		\begin{center}		
			\begin{tabular}{ll}
				\hline
				hyperparameter & value \\ \hline
				number of hidden layers 	& 	3 \\
				neurons per hidden layer   	& 	64 	\\
				dropout probability & 	0.2 		\\
				activation f. hidden layer  & 	LeakyReLU,\\& $\alpha =0.01$ \\
				initialization of hidden layer & Xavier uniform,\\& $\sim U(-0.5,0.5)$ \\
				activation f. output layer A/C  	& 	linear 	\\
				initial weights all layers A/C   & $\sim U(-1,1)$ \\
				optimizer  	&  	Adam, $\beta_1=0.9$, \\&$\beta_2=0.999$, no gradient \\&clipping, decay, fuzz \\&factor or AMSGrad \\
				error metric & Huber loss \\
				target network update frequency $m$ 	&  	$4000$ \\ \hline
			\end{tabular}
		\end{center}
	\end{table}

	\begin{table}[h!]\captionsetup{width=\columnwidth}\caption{Hyperparameters of the algorithm}\label{tab:RL}
		\begin{center}		
			\begin{tabular}{ll}
				\hline
				hyperparameter & value \\ \hline
				discount factor $\gamma$  	& 	$0.999$ 	\\
				$\xi_{\text{temp}}$& \SI{0}{}\tdd{ZAHL ÜBERPRÜFEN} \\
				ERM size $M$ 	& 	$\SI{1e5}{}$\\
				macro-batch size $B$ 	& 	256 \\
				mini-batch size $t$   	& 	80 	\\
				$\alpha$ learning rate & 	\SI{5e-4}{} 		\\
				$\sigma$ learning rate & 	\SI{2e-4}{} 		\\
				$\beta$ learning rate & 	\SI{5e-5}{} 		\\
				$\lambda_{\text{high}}$ &	0.8	\\
				$\lambda_{\text{low}}$	&	0.2	\\	
				exploration &  	$\varepsilon_{\text{min}}=0.01$, decay rate $\varpi= 0.999$ \\
				 \hline
			\end{tabular}
		\end{center}
	\end{table}

\end{document}